\documentclass[10pt,conference]{IEEEtran}
\usepackage{cite}
\usepackage{amsmath,amssymb,amsfonts}
\usepackage{algorithmic}
\usepackage{graphicx}
\usepackage{textcomp}
\usepackage{xcolor}
\def\BibTeX{{\rm B\kern-.05em{\sc i\kern-.025em b}\kern-.08em
    T\kern-.1667em\lower.7ex\hbox{E}\kern-.125emX}}

\usepackage{booktabs} % \toprule, \midrule and \bottomrule, for prettier tables
\usepackage{tabularx} % 'X' column type, spans remaining table space
\usepackage{balance} % \balance to balance column height
\usepackage{multirow} % Merge vertical cells, https://tex.stackexchange.com/questions/72945
\usepackage{siunitx} % 'S' column type, which aligns on decimal point
\usepackage{hyperref}
\usepackage{graphicx,cleveref}
\newcommand{\boxbox}[2]{
    \begin{center}\fbox{\parbox{#1\columnwidth}{#2}}\end{center}
}

\IEEEoverridecommandlockouts
\usepackage{tikz}
\newcommand\copyrighttext{%
  \footnotesize Copyright \textcopyright 2019 IEEE. Personal use of this material is permitted.
  Permission from IEEE must be obtained for all other uses, in any current or future 
  media, including reprinting/republishing this material for advertising or promotional 
  purposes, creating new collective works, for resale or redistribution to servers or 
  lists, or reuse of any copyrighted component of this work in other works. 
  }
\newcommand\copyrightnotice{%
\begin{tikzpicture}[remember picture,overlay]
\node[anchor=south,yshift=10pt] at (current page.south) {\fbox{\parbox{\dimexpr\textwidth-\fboxsep-\fboxrule\relax}{\copyrighttext}}};
\end{tikzpicture}%
}

\begin{document}

\title{An Additional Set of (Automated) Eyes:\\Chatbots for Agile Retrospectives}

\author{\IEEEauthorblockN{Christoph Matthies, Franziska Dobrigkeit, Guenter Hesse}
\IEEEauthorblockA{\textit{Hasso Plattner Institute, University of Potsdam, Germany} \\
christoph.matthies@hpi.de, franziska.dobrigkeit@hpi.de, guenter.hesse@hpi.de}
}

\maketitle

% Render IEEE copyright notice
\copyrightnotice

\begin{abstract}
Recent advances in natural-language processing and data analysis allow software bots to become virtual team members, providing an additional set of automated eyes and additional perspectives for informing and supporting teamwork.
In this paper, we propose employing chatbots in the domain of software development with a focus on supporting analyses and measurements of teams' project data.
The software project artifacts produced by agile teams during regular development activities, e.g. commits in a version control system, represent detailed information on how a team works and collaborates.
Analyses of this data are especially relevant for agile retrospective meetings, where adaptations and improvements to the executed development process are discussed.
Development teams can use these measurements to track the progress of identified improvement actions over development iterations.
Chatbots provide a convenient user interface for interacting with the outcomes of retrospectives and the associated measurements in a chat-based channel that is already being employed by team members.
\end{abstract}


\section{Introduction}
Software tools that help software users and software developers in their daily tasks have been created for as long as code has been written~\cite{Paikari2018}.
Today, a countless number of such development support tools exist, from large and powerful to small and simple. 
They assist in various activities such as architectural design, test case generation~\cite{Wasserman1990} or collaboration within teams~\cite{Lanubile2010}.

\subsection{Chatbot Definition}
Traditionally, support tools which take actions on behalf of a user are referred to as \emph{software agents}~\cite{Nwana1996}, from the Latin ``agere'', meaning ``to take action''.
The importance of automation and autonomy involved with many of these systems is highlighted in the designation \emph{softbots}~\cite{hedberg1995intelligent}, short for software robots.
More recently, the terms \emph{chatbot}, \emph{chatterbot} or simply \emph{bot} have been used to refer to support tools that employ a conversational-style user interface~\cite{Lebeuf2018}.
These types of text interfaces to software services have gained popularity due to the increasing role of instant messaging in the workplace as well as in social encounters~\cite{Chan2005}.
Messaging platforms which are widely adopted include \emph{Facebook Messenger} for social networking or \emph{Slack} for work-related communication~\cite{Perkel2014}.
Therefore, these platforms, where users already frequently interact and collaborate, are where chatbots are often employed~\cite{Lin2016}.

\subsection{Chatbot Operation}
\begin{figure}[htb]
    \centering
    \includegraphics[width=0.75\columnwidth]{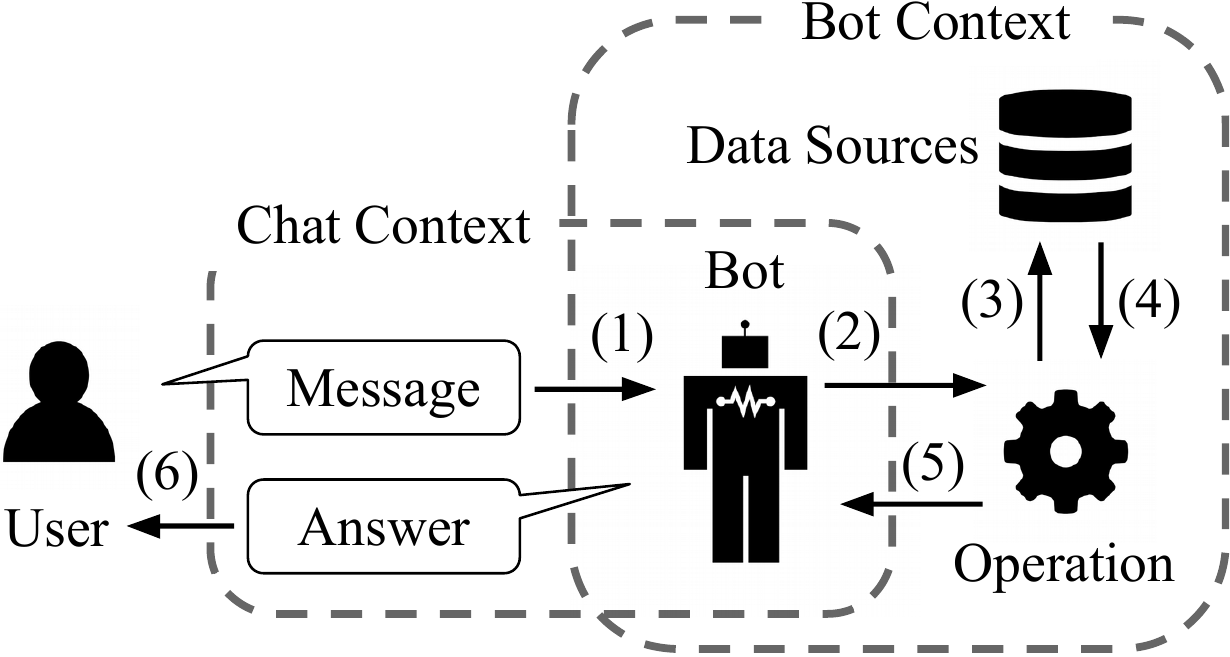}
    \caption{Overview of Chatbot operations, adapted from Inokuchi et al.~\cite{Inokuchi2016}}
    \label{fig:chatops}
\end{figure}

At its simplest, a (chat)bot is a computer program which performs various predetermined operations, receiving commands via chat messages and performing the requested action~\cite{Inokuchi2016}.
\Cref{fig:chatops} depicts the general interaction flow through such a system.
Initially, a user sends a command as a chat message to the bot (1).
The bot parses the message and performs operations based on the message's contents (2).
To fulfill the request, the bot might need to request (3) and receive (4) additional information from third-party sources, such as knowledge bases or internet resources~\cite{Paikari2018}.
Finally, the bot computes a result (5), compiles a chat message and sends it as a reply to the user (6).

In this manner chatbots can provide a conduit between users and tools, integrating external services and additional information sources into already existent communication channels~\cite{Lebeuf2018}.

\subsection{Chatbots in Agile Teams}
With the improvement of natural-language processing as well as big data analytics and their application in conversational bots~\cite{Lebeuf2018}, bots can grow from simple scripted tools to ``virtual teammates''\cite{Lebeuf2017}, informing and supporting teamwork.
\boxbox{0.9}{In this paper, we propose an approach for employing chatbots in teams with a focus on analyses and measurements of the data produced by the team, particularly in the domain of software development.}
Modern software development teams employing agile methods spend a substantial amount of their time interacting with development tools, such as \emph{Version Control Systems} (VCS), during regular development activities.
The software project artifacts~\cite{Fernandez2018} produced using these systems, i.e. commits in a VCS containing code changes, represent a ``gold-mine of actionable information''~\cite{Guo2016}, i.e. knowledge on how a team works and collaborates~\cite{Rosen:2015:FSE, Matthiesb, Santos2016}.
While analyses of this data can be used for a variety of purposes, they are especially relevant for software development teams in the context of their efforts to adapt and improve the executed development process.
For example, relevant metrics such as burndown charts can be constructed by evaluating the software project artifacts of a development iteration~\cite{Derby2006}.
Chatbots provide a convenient user interface for interacting with these types of analyses~\cite{Lebeuf2018} in a chat-based channel already being used by teams on a daily basis.

\section{Related Work on Chatbots}
A wide variety of application scenarios for chatbot usage in teams has been described, ranging from information processing and sharing to detecting and monitoring activities in team communications and even providing recommendations regarding possible next tasks~\cite{Lebeuf2018}.
Of course, they have also been used to search for and send animated cat gifs~\cite{Eppink2014} to team members~\cite{Lin2016}.

\subsection{Software Engineering}
Software developers belong to the early adopters of automation and bot usage as they are familiar with automated tools increasing code quality and team productivity~\cite{Lebeuf2018}.
A particularly interesting approach is the \emph{BuildBot} by Ablett et al., a chatbot in the true nature of the term: it communicates with developers using sound~\cite{Ablett2007}.
The bot uses a robotic interface, a small Sony AIBO robot dog, which monitors the build status of software within an agile continuous integration approach.
Should the build be broken, e.g. because tests fail, the robot walks to the developer whose code is responsible for the failure and notifies them (in a playful way).
The authors argue that the system increased awareness of the software's build status and helped agile teams with self-supervision~\cite{Ablett2007}.
In modern software development, chat solutions have disrupted previous software development processes and have replaced email in some cases~\cite{Lin2016, Kafer2018}.
Fitzpatrick et al. combined the informal discussions developers have about code with version control information reporting code changes in a ticker tape form~\cite{Fitzpatrick2006}.
They point out that the studied software developers used the close integration of versioning information with chat functionality for such varied tasks as growing team culture, marking phases of work or managing work interruptions.
Similarly, Lebeuf et al. report on industry teams that use chatbots to provide instructions for development procedures, e.g. merging feature branches, to monitor website outages, or to manage code deployments~\cite{Lebeuf2017, Lebeuf2018}.

\subsection{Human Factors}
In addition to uses in communication and data access, chatbots have previously also been employed to tackle issues related to human factors in software engineering.
By sending commands to bots via chat messages in group chats, every operation is shared with the team as well as logged and persisted in the chat log~\cite{Inokuchi2016}.
This enables transparency and awareness of other team members interactions with the chatbot and its functions.
It enables nontechnical team members to engage with the bot's capabilities without explicitly needing domain expertise~\cite{Lebeuf2018}.
Having human team members and bots share the same chat context enables use cases such as adopting the chat solution as a distributed command line or a collaborative debugger~\cite{Chan2005}.
Especially relevant to the topic of employing chatbots for process improvement approaches within agile teams is the idea of using bots to regulate individual and team tasks and goals~\cite{Storey2016}.
For collaboration to succeed in a team, all members must understand and share the goals set for the team as well as the actions necessary to achieve these goals~\cite{Lebeuf2017}.
Bots can initiate and track reminders set in earlier meetings as well as help to monitor and visualize progress towards certain team goals~\cite{Storey2016}.

\section{Chatbots in Agile Retrospectives}
In order to improve the executed software development process within a team, the current status has to be measured, so that changes in the future can be detected.
The agile process framework currently most popular in industry~\cite{ScrumAlliance2017}, \emph{Scrum}, calls for a specific meeting with the goal of process improvement: the \emph{retrospective meeting}~\cite{Schwaber2017}.

\subsection{The Retrospective Meeting}
As the name suggests, in the dedicated retrospective meeting the team looks back at the most recent development iteration and decides which aspects of the process should be kept and what should be changed in the future.
Action items, i.e. issues that should be improved are recorded as outcomes of the meeting.
In the next retrospective, the development team then decides whether headway has been made on the previously defined actions items~\cite{Kniberg2007}.
While this decision relies on the team members understanding of their executed process, the official Scrum Guide also specifies that ``decisions to optimize value and control risk are made based on the perceived state of the artifacts.''
It goes on to define a task which is to ``detect incomplete transparency by inspecting the artifacts, sensing patterns [...] and detecting differences between expected and real results''~\cite{Schwaber2017}.
Similarly, Derby and Larsen suggest to ``start with the hard data'' in retrospectives~\cite{Derby2006}, including metrics such as velocity, defect count, number of stories completed or amount of refactored code.

\begin{figure}[htb]
    \centering
    \includegraphics[width=\columnwidth]{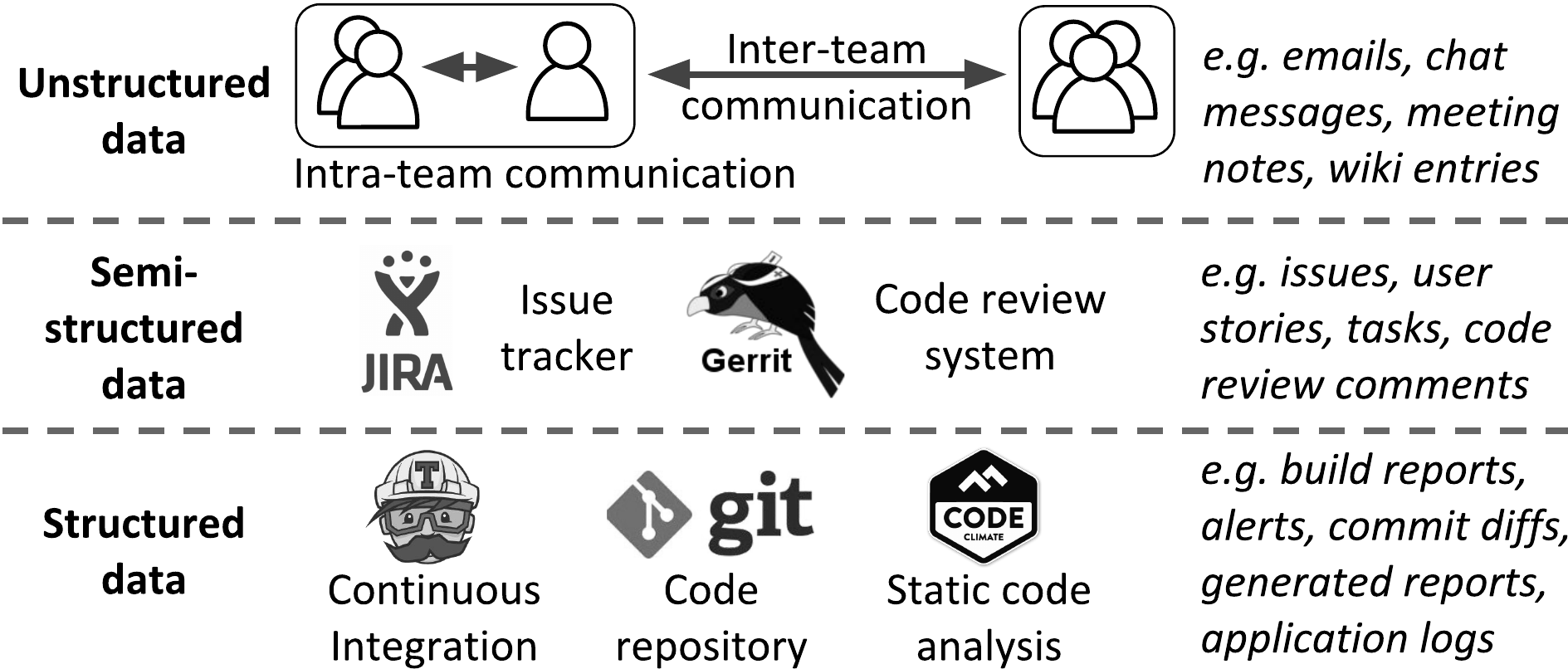}
    \caption{Sources and examples of project artifacts in the software development domain providing data of different levels of structure.}
    \label{fig:data_sources}
\end{figure}

\Cref{fig:data_sources} provides an exemplary overview of the different sources of development artifacts available in an agile software development team.
In teams with large code bases and therefore a large number of development artifacts to inspect, collection and analysis is a challenge, especially for human actors.
However, a bot, equipped with human instruction on what patterns to detect, e.g. desired outcomes from retrospective action items, can analyze large quantities of data.

\subsection{Chatbots for Process Improvement}
Developers have built or customized chatbots to support their daily lives, also outside of work and development activities.
Early examples of this are bots which have been used to assist with information retrieval~\cite{Voorhees1994}.
More recent examples include bots that can help decide where to go to lunch or help keep a grocery list as well as provide entertainment, i.e. to ``search for images and to then add moustaches to them''~\cite{Lin2016}.
With the varied background of bot tasks in mind, we propose having a bot track the progress of retrospective action items, which can come from a variety of contexts.
As agile software development teams already spend significant amounts of time communicating in chat solutions and sometimes also already use these during retrospectives~\cite{Standuply2019}, this means no context switch is necessary.
Instead of the Scrum Master or an agile coach analyzing project data and providing a different perspective to team members during retrospectives, it could be a bot that retrieves and processes the information, leaving humans more time to interpret it.

\subsection{Integration into the Agile Processes}
\begin{figure}[htb]
    \centering
    \includegraphics[width=0.85\columnwidth]{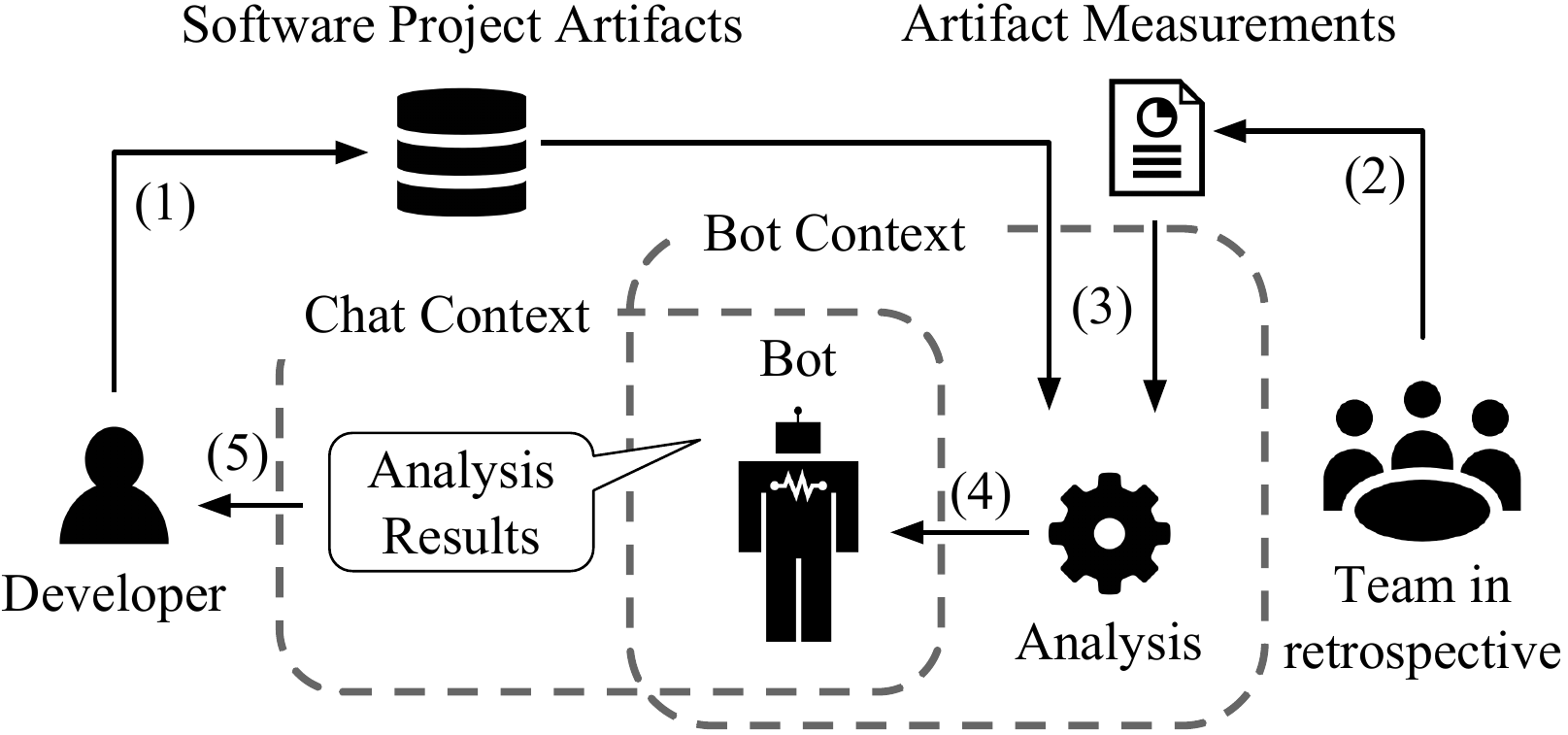}
    \caption{Overview of using a chatbot for analysis of software project artifacts}
    \label{fig:vision}
\end{figure}
\Cref{fig:vision} visualizes the operation of the envisioned chatbot and its integration into the agile process framework.
Agile software developers produce project artifacts in their daily activities during a development iteration (1).
At the end of an iteration, they hold a retrospective meeting, focusing on those practices that went well and should be continued as well as defining action items for those that should be changed.
To track these action items, measurements based on the project artifacts are created by the team (2), e.g. the number of commits into the VCS which increase code complexity but do not provide tests.
The bot applies these measurements to the collected project data and stores the analysis values over time (4).
Using the chat service, it communicates the results, i.e. the change of the measurement, to the development team members (5).
This information can be used in the next retrospective to initiate discussions on the state of an action item, on the basis of concrete data points, in addition to---and possibly in contrast to---the perceptions of team members.

\section{Integration into Existing Tools}
The use of tools is vital to collaboration within modern software development teams, especially for geographically distributed teams.
Tools enable the automation, and control of the entire development process~\cite{Lanubile2010}.
Automation is also key to many common practices of modern agile software development, such as frequent product deliveries, which would otherwise not be feasible~\cite{Ebert2016}.
There is a range of tools available, specifically aimed at supporting the retrospective meetings of agile teams through automation, by setting reminders, archiving action items~\cite{GoReflect2019}, or facilitating activities~\cite{Retrium2019}.
The chat solution Slack, which is popular for work-related instant messaging~\cite{Lin2016},
features extension points and APIs for third-party applications and bots to interact with users via a conversation~\cite{SlackBotAPI}.
These possibilities have already been used to create initial chatbots on the Slack platform that support agile teams in their retrospectives~\cite{Standuply2019, Sharp2019, McAuliffe2019}.
These bots remind team members of the retrospective meeting and can record the individual statements of developers, summarizing results and archiving outcomes.
While they automate the process, this approach still fully relies on team members' perceptions to provide the inputs for discussion.
Currently available bots can automate the tedious organizational tasks, but do not provide an additional perspective based on the available project data.
In order to provide this additional perspective, the chatbot must be enabled to assess a team's situation by measuring the team's development data.
This can be achieved through various metrics designed for agile practices~\cite{Ju2018, Matthies2016,Perkusich2016} or by using tools, such as the git command line~\cite{GitDocumentation}, that developers are already familiar with from their daily development activities.

\begin{figure}[htb]
    \centering
    \includegraphics[width=1\columnwidth]{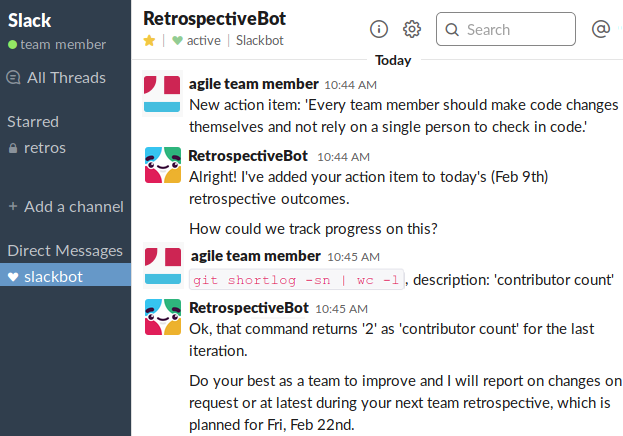}
    \caption{User interface mockup of the interaction between a developer and a retrospective bot tasked with tracking retrospective action items.}
    \label{fig:mockup}
\end{figure}

\Cref{fig:mockup} shows an exemplary interaction during a retrospective between a development team member and the envisioned retrospective bot.
A team member has identified an improvement for the next development iteration, i.e. that everyone should check in code.
They notify the retrospective bot of this new action item and provide a short command line call, measuring the number of unique contributors in the VCS.
This measurement acts as a proxy for assessing progress on the new action item.
At this point, the bot takes over, repeatedly taking measurements using the provided code statement, and will inform team members of the status of the action item.
In the next retrospective the results can then be discussed and interpreted by the team, as a basis for data-informed improvements.

\section{Conclusion}
It seems certain that software developers will see more automation and bots being introduced to support their workflow and development-related activities~\cite{Lebeuf2018}.
This includes both coding activities as well as tending to and improving the development process executed in teams.
This proposal for using chatbots in agile retrospectives as an additional set of automated eyes on the software project data of teams is a step in this direction.

\bibliographystyle{IEEEtran}
\bibliography{minified}

\end{document}